\documentclass[journal,10pt]{IEEEtran}

\usepackage{amsmath,amssymb,amsthm}
\usepackage{cite}
\usepackage{microtype}
\usepackage{float}
\usepackage{graphicx,tikz}
\usetikzlibrary{arrows.meta,calc}

\newtheorem{assumption}{Assumption}
\newtheorem{lemma}{Lemma}
\newtheorem{theorem}{Theorem}
\newtheorem{proposition}{Proposition}
\newtheorem{corollary}{Corollary}
\newtheorem{remark}{Remark}

\newcommand{\R}{\mathbb{R}}
\newcommand{\RH}{\mathcal{RH}_{\infty}}
\newcommand{\Linfty}{\mathcal{L}_{\infty}}
\newcommand{\Lone}{\mathcal{L}_{1}}
\newcommand{\norm}[1]{\left\lVert #1\right\rVert}
\newcommand{\abs}[1]{\left\lvert #1\right\rvert}
\newcommand{\dd}{\mathrm{d}}
\newcommand{\rdeg}{\operatorname{rdeg}}
\newcommand{\col}{\operatorname{col}}

\title{Funnel-Precompensated Disturbance {Observer} for Uncertain Minimum-Phase SISO Linear Systems of Arbitrary Relative Degree}

\author{Hyuntae Kim and Thomas Berger%
\thanks{H. Kim is with the Department of Mechanical Engineering, University of Maryland, College Park, MD 20742, USA (e-mail: htkim26@umd.edu).}%
\thanks{T. Berger is with the Institute of Mathematics, Martin Luther University Halle-Wittenberg, Theodor-Lieser-Str. 5, 06120 Halle (Saale), Germany (e-mail: thomas.berger@mathematik.uni-halle.de). Thomas Berger acknowledges funding by the Deutsche Forschungsgemeinschaft (DFG, German Research Foundation) -- Project-ID~544702565.}%
}

\begin{document}
\maketitle
\thispagestyle{empty}
\pagestyle{empty}

\begin{abstract}
Disturbance observers (DOBs) recover nominal behavior, but fast filters risk peaking. We propose a funnel-precompensated DOB for uncertain minimum-phase single-input single-output linear systems of known relative degree. A stable-factor residual yields a minimum-phase channel. At relative degree two, gain conditions require positive plant high-frequency gains and one positive common design gain; for higher degrees, the common-design result also uses a plant-gain interval and mismatch condition. Global boundedness, practical nominal-output recovery, and open-funnel initialization are established.
\end{abstract}


\section{Introduction}
Disturbance-observer-based (DOB) control augments a nominally stabilizing controller with an inner loop that compensates matched disturbances and model mismatch \cite{shim2016tutorial,sariyildiz2020overview}. Conventional realizations use a low-pass $Q$-filter to make the nominal inverse proper; reducing its time scale improves recovery over an admissible uncertainty class but can introduce fast-dynamics peaking \cite{shim2016tutorial}.

Funnel control replaces a fixed high gain by an error-dependent gain near a prescribed boundary. {A funnel pre-compensator (FPC) provides an artificial output and its derivatives \cite{berger2018fpc,berger2018tac}.} For systems of arbitrary relative degree, FPC cascades preserve minimum phase and the original relative degree \cite{lanza2022output}. High-gain-observer funnel control handles arbitrary relative degree through a sufficiently large fixed observer gain \cite{chowdhury2019hgo}. The methods in \cite{hackl2013two,schaa2026input} target direct tracking; the present work regulates an uncertain DOB residual.

Stable coprime-factor graph representations describe nominal input--output behavior \cite{vidyasagar1985}. We connect arbitrary-relative-degree FPC theory to DOB design through a stable-factor residual. {{The resulting minimum-phase channel preserves each plant's relative degree and positive high-frequency gain without requiring nominal stabilization of the actual plant. At relative degree two, plant high-frequency gains and the common design gain need only be positive; from relative degree three, the common-design result also uses a plant-gain interval and mismatch condition.}} The residual realization shows that the bounded equivalent disturbance enters through the matched input channel required by the FPC result. A stable recovery identity gives nominal-output bounds. Finite-horizon observability bounds the plant and controller realization states and completes the implemented-state boundedness argument. For bounded reciprocal funnels, an admissible stable weighting gives a filtered-estimation-error bound. Open funnels remove residual-channel and FPC initial-domain restrictions. Online implementation uses the residual and, for relative degree at least two, the FPC states.

The numerical study examines open-profile initialization, a fixed-$Q$-DOB comparison with matched sampled correction $L_2$ norm, finite-profile rejection, fixed-design uncertainty, and relative degrees two through four.

{\emph{Notation:} $s$ is the Laplace variable, $I_k$ the $k\times k$ identity, and $\rdeg H$ the relative degree of rational $H$. Set $\R_{\geq0}:=[0,\infty)$. The sets $\R[s]$, $\RH$, $\Linfty$, and $\Lone$ denote real polynomials, stable proper real-rational transfers, essentially bounded signals, and absolutely integrable signals on $\R_{\geq0}$, respectively; $C^k$ denotes $k$-times continuously differentiable functions. For $H\in\RH$ with impulse response $h=h_0\delta_{\mathrm D}+\bar h$, where $h_0\in\R$, $\bar h\in\Lone$, and $\delta_{\mathrm D}$ is the Dirac impulse, define $\norm{H}_{\Lone}:=\abs{h_0}+\norm{\bar h}_{\Lone}$. Signal infinity norms are essential suprema, and unsubscripted vector norms are Euclidean. Each implemented proper rational operator has a fixed minimal realization. The expression $H(s)x$ denotes the output of the causal linear time-invariant (LTI) operator with transfer function $H$ driven by $x$ under the stated initial condition. Transfer identities are causal convolution identities; unless stated otherwise, the associated plant, controller, and filter realizations start at zero. A Carath\'eodory solution is absolutely continuous and satisfies the state equations almost everywhere. For scalar funnel function $f$, write $f(t)^{-1}=1/f(t)$ where $f(t)>0$. Set $\col(a_1,\ldots,a_m):=[a_1^{\top},\ldots,a_m^{\top}]^{\top}$. Landau symbols refer to $t\downarrow0$.}

\section{Problem Formulation and DOB Residual}
Let $\Theta$ be an index set and consider the uncertain single-input single-output (SISO) plant
\begin{equation}
 P_{\theta}(s)=\frac{B_{\theta}(s)}{A_{\theta}(s)},\qquad
 y=P_{\theta}(u+d),\qquad \theta\in\Theta,
 \label{eq:plant}
\end{equation}
where $A_\theta,B_\theta\in\R[s]$, $u$ is the applied input, $y$ is the measured output, and $d\in\Linfty$ is a matched disturbance. The actual index $\theta$ is fixed but unknown to the designer. All closed-loop signals may depend on this fixed index; the dependence is suppressed unless needed explicitly. The nominal plant and the predesigned proper controller are
\begin{equation}
 P_n=\frac{B_n}{A_n},\qquad C=\frac{N_c}{D_c},\qquad
 u_c=C(w-y),
 \label{eq:nominal_controller}
\end{equation}
where $A_n,B_n,N_c,D_c\in\R[s]$, $w\in\Linfty$ is the reference signal, and $u_c$ is the nominal-controller output. The compensation input is inserted as
\begin{equation}
 u=u_c+v=u_c-\hat d_F,\qquad \hat d_F:=-v.
 \label{eq:input}
\end{equation}
We use $v$ for the compensation signal applied to the plant and write $\hat d_F=-v$ for the conventional disturbance-estimate interpretation. The objective is to construct $v$ from an implementable residual built from the nominal input--output relation so that, for every fixed unknown $\theta\in\Theta$, the closed loop is globally well posed and bounded and admits tunable recovery of the nominal closed-loop output, without requiring $C$ to stabilize $P_\theta$.

\begin{assumption}\label{ass:plants}
For every $\theta\in\Theta$, $A_{\theta}$ is monic, $(A_{\theta},B_{\theta})$ is coprime, and $P_{\theta}$ is strictly proper and minimum phase, meaning that $B_\theta$ is Hurwitz. All plants have the same known relative degree $r\geq1$. {{For every $\theta\in\Theta$, define the high-frequency gain by
\[
 \gamma_\theta:=\lim_{s\to\infty}s^rP_{\theta}(s)
\]
and assume $\gamma_\theta>0$.
When $r\geq3$, assume additionally that these gains lie in a known interval
\[
 0<\underline\gamma\leq\gamma_\theta\leq\overline\gamma<\infty.
\]}}
Consequently, $\deg B_\theta=\deg A_\theta-r$, and the leading coefficient of $B_\theta$ is $\gamma_\theta$.
\end{assumption}

\begin{assumption}\label{ass:nominal_model}
{The nominal model introduced in \eqref{eq:nominal_controller} is $P_n=P_{\theta_n}$ for some $\theta_n\in\Theta$, and we take $A_n:=A_{\theta_n}$ and $B_n:=B_{\theta_n}$.}
\end{assumption}

\begin{assumption}\label{ass:nominal_loop}
The pair $(D_c,N_c)$ is coprime, $D_c$ is monic, and $C$ internally stabilizes $P_n$. Equivalently,
\[
 \chi_n:=A_nD_c+B_nN_c
\]
is Hurwitz.
\end{assumption}

Let $n_n:=\deg A_n$ and choose a monic Hurwitz polynomial $\Lambda$ of degree $n_n$. Define the stable factors
\begin{equation}
 M_n:=\frac{A_n}{\Lambda},\qquad
 N_n:=\frac{B_n}{\Lambda},\qquad P_n=N_nM_n^{-1}.
\end{equation}
The proposed residual is
\begin{equation}
 e_F:=M_ny-N_nu_c
 =y+\frac{A_n-\Lambda}{\Lambda}y-\frac{B_n}{\Lambda}u_c.
 \label{eq:residual}
\end{equation}
{The transfer functions $(A_n-\Lambda)/\Lambda$ and $B_n/\Lambda$ are stable and strictly proper; hence the residual is implementable without differentiating the measured output.} Define
\begin{equation}
 \zeta_F:=M_ny-N_nu.
\end{equation}
Using \eqref{eq:input},
\begin{equation}
 e_F=\zeta_F-N_n\hat d_F.
\end{equation}
Interpreting $N_n\hat d_F$ as an estimate of the implementable filtered lumped-disturbance signal $\zeta_F$, this identity gives $e_F$ as the corresponding estimation error. For the nominal plant $P_\theta=P_n$, $\zeta_F=N_nd$ and
\begin{equation}
 e_F=N_n(d-\hat d_F).
\end{equation}

For comparison, let $\tau>0$ be a filter time-scale parameter and let $Q_{\tau}\in\RH$ satisfy $Q_{\tau}(0)=1$ and $Q_{\tau}P_n^{-1}$ be proper. A conventional DOB may be written as
\begin{equation}
 \hat d_Q=Q_{\tau}(P_n^{-1}y-u),\qquad u=u_c-\hat d_Q,
\end{equation}
which implies
\begin{equation}
 (1-Q_{\tau})\hat d_Q=Q_{\tau}(P_n^{-1}y-u_c).
\end{equation}
Both architectures therefore use proper filters of the same nominal-inverse discrepancy $P_n^{-1}y-u_c$. In particular,
\[
 e_F=N_n(P_n^{-1}y-u_c),
\]
so the proposed residual realizes this weighting without evaluating the improper inverse $P_n^{-1}$ directly.

\section{Residual-Channel Structure}
Define
\begin{equation}
 \chi_{\theta}:=A_{\theta}D_c+B_{\theta}N_c.
\end{equation}

\begin{lemma}[Residual channel]\label{lem:channel}
Under Assumptions~\ref{ass:plants}, \ref{ass:nominal_model}, and \ref{ass:nominal_loop}, and with zero initial states in the plant, nominal controller, and the realizations implementing $M_n$ and $N_n$, the residual satisfies
\begin{equation}
 e_F=G_{\theta}(v+\bar d_{\theta}),\qquad
 G_{\theta}:=\frac{B_{\theta}\chi_n}{\Lambda\chi_{\theta}},
 \label{eq:error_channel}
\end{equation}
where
\begin{equation}
 \bar d_{\theta}:=d+\Delta_{\theta}w,\qquad
 \Delta_{\theta}:=\frac{N_c(B_{\theta}A_n-B_nA_{\theta})}{B_{\theta}\chi_n}.
 \label{eq:equiv_dist}
\end{equation}
The transfer $\Delta_{\theta}$ is stable and proper. Moreover, $\rdeg G_{\theta}=r$, {its high-frequency gain is $\gamma_\theta$}, and its finite zeros are stable. The polynomial $\chi_{\theta}$ need not be Hurwitz.
\end{lemma}

\begin{proof}
From \eqref{eq:plant}--\eqref{eq:input},
\[
 y=\frac{P_{\theta}C}{1+P_{\theta}C}w
 +\frac{P_{\theta}}{1+P_{\theta}C}(v+d).
\]
Since $e_F=M_n(1+P_nC)y-N_nCw$, substitution gives
\[
 G_{\theta}=M_nP_{\theta}\frac{1+P_nC}{1+P_{\theta}C},\qquad
 \Delta_{\theta}=\frac{C(P_{\theta}-P_n)}{P_{\theta}(1+P_nC)}.
\]
The polynomial descriptions reduce these expressions to \eqref{eq:error_channel}--\eqref{eq:equiv_dist}. Any finite pole of $\Delta_{\theta}$ is among the roots of $B_{\theta}\chi_n$ and is therefore stable. {If $N_c\equiv0$, then $\Delta_{\theta}\equiv0$. Otherwise, let $n_{\theta}:=\deg A_{\theta}$, $m_c:=\deg D_c$, and $k_c:=\deg N_c\leq m_c$. Since all plants have relative degree $r$,
\[
 \deg(B_\theta A_n-B_nA_\theta)\leq n_\theta+n_n-r.
\]
Thus the numerator degree in \eqref{eq:equiv_dist} is at most $n_\theta+n_n-r+k_c$, while the denominator degree is $n_\theta+n_n-r+m_c$. Hence $\Delta_\theta$ is proper.}

The numerator and denominator degrees of $G_{\theta}$ differ by $r$, and {$\lim_{s\to\infty}s^rG_{\theta}(s)=\gamma_\theta$}. After any cancellations, the finite zeros are a subset of the roots of $B_{\theta}\chi_n$, all of which are stable. Uncancelled roots of $\chi_{\theta}$ are poles of $G_{\theta}$, not finite zeros.
\end{proof}

The factor $M_n=A_n/\Lambda$ is biproper, meaning that it has relative degree zero, and therefore does not add relative degree; Lemma~\ref{lem:channel} gives $\rdeg G_\theta=r$. A strictly proper residual premultiplier would add its own relative degree and require a longer FPC cascade.

Because $\Delta_{\theta}\in\RH$, the additive signal $\bar d_{\theta}$ is bounded. For application of the FPC result, choose a minimal realization of the induced residual channel $G_\theta$,

\begin{equation}
 \dot x_R=A_Rx_R+B_R(v+\bar d_{\theta}),\qquad
 e_F=C_Rx_R,
 \label{eq:residual_realization}
\end{equation}

where $n_R$ is the minimal-realization order (the McMillan degree) of $G_\theta$, $A_R\in\R^{n_R\times n_R}$, $B_R\in\R^{n_R}$, $C_R\in\R^{1\times n_R}$, and $x_R\in\R^{n_R}$. This state-space realization is used only for analysis and supplies the form required by the FPC theorem. Because the finite zeros of $G_\theta$ are stable, this realization has asymptotically stable zero dynamics and belongs to the linear subclass of the arbitrary-relative-degree FPC framework in \cite[Rems.~3.3 and 3.7]{lanza2022output}. {The relative degree and high-frequency gain give}
{
\begin{equation}
 \begin{aligned}
 C_RA_R^kB_R&=0,&& k=0,\ldots,r-2\ \text{ when }r\geq2,\\
 C_RA_R^{r-1}B_R&=\gamma_\theta.
 \end{aligned}
 \label{eq:markov_parameters}
\end{equation}
Consequently, for $r\geq2$,
\[
 C_RA_R^kB_R\bar d_\theta(t)=0,
 \qquad k=0,\ldots,r-2,
\]
and, for every $r\geq1$,
\[
 C_RA_R^{r-1}B_R\bar d_\theta
 =\gamma_\theta\bar d_\theta\in\Linfty.
\]}
Thus $B_R\bar d_\theta$ satisfies the disturbance condition of the cited linear subclass without requiring derivatives of $w$ or $d$.

For analysis only, define the nominal closed-loop output by
\begin{equation}
 y_0:=\frac{P_nC}{1+P_nC}w.
 \label{eq:y0}
\end{equation}
With zero initial states for the plant, controller, and stable-factor realizations, subtracting the nominal relation from \eqref{eq:residual} gives
\begin{equation}
 y-y_0=H_ye_F,
 \qquad H_y:=(M_n+N_nC)^{-1}=\frac{\Lambda D_c}{\chi_n}\in\RH.
 \label{eq:npr_map}
\end{equation}

\section{Funnel-Precompensated DOB}
For each integer $k\geq1$, define $\Phi_k$ as the set of all $\varphi\in C^k(\R_{\geq0},\R_{\geq0})$ satisfying $\varphi^{(j)}\in\Linfty$, $j=0,\ldots,k$, $\varphi(t)>0$ for $t>0$, and $\liminf_{t\to\infty}\varphi(t)>0$. The value at $t=0$ may be zero. {{For $r\geq2$, choose $\varphi_1\in\Phi_r$. When $r\geq3$, additionally choose $\varphi_p\in\Phi_r$ such that $\varphi_p=\rho\varphi_1$ for some $\rho>1$.}}

For $r\geq2$, select $a_1,\ldots,a_r>0$ so that
\[
 A_f:=\begin{bmatrix}
 -a_1&1&&0\\
 -a_2&0&\ddots&\\
 \vdots&\vdots&\ddots&1\\
 -a_r&0&\cdots&0
 \end{bmatrix}
\]
is a Hurwitz matrix in $\R^{r\times r}$. For $Q_f=Q_f^{\top}\in\R^{r\times r}$ with $Q_f\succ0$, let $X_f=X_f^{\top}\in\R^{r\times r}$ with $X_f\succ0$ be the unique solution of
\begin{equation}
 A_f^{\top}X_f+X_fA_f=-Q_f.
 \label{eq:Af}
\end{equation}
Partition
$X_f=\left[\begin{smallmatrix}X_{11}&X_{12}\\X_{12}^{\top}&X_{22}\end{smallmatrix}\right]$,
where $X_{11}\in\R$, $X_{12}\in\R^{1\times(r-1)}$, and $X_{22}\in\R^{(r-1)\times(r-1)}$. Since $X_f\succ0$, its principal submatrix $X_{22}\succ0$, so $X_{22}^{-1}$ is well defined. Define
\begin{equation}
 \mathbf p:=
 \begin{bmatrix}
  1\\
  -X_{22}^{-1}X_{12}^{\top}
 \end{bmatrix}
 =:[p_1,\ldots,p_r]^{\top}.
 \label{eq:p}
\end{equation}
A constructive feasible family takes $a_j:=\binom{r}{j}\lambda_f^j$, $j=1,\ldots,r$, from $(s+\lambda_f)^r$ with $\lambda_f>0$, and $Q_f:=I_r$; the corresponding Lyapunov solution determines $\mathbf p$ through \eqref{eq:p} \cite[Rem.~3.4]{lanza2022output}. {{For $r=2$, choose one common $\widetilde\gamma>0$; \cite[Thm.~2]{berger2018fpc} imposes no condition linking it to the positive plant gain $\gamma_\theta$. For $r\geq3$, choose one $\widetilde\gamma>0$ and define}}
{
\begin{align}
 \mu_\gamma
 &:=\sup_{\theta\in\Theta}
 \left|1-\frac{\gamma_\theta}{\widetilde\gamma}\right|,
 \nonumber\\
 \bar\mu_{r,\rho}
 &:=\min\!\left\{\frac{\rho-1}{r-2},
 \frac{\rho}{4\rho^2(\rho+1)^{r-2}-1}\right\},
 \nonumber\\
 \mu_\gamma&<\bar\mu_{r,\rho}.
 \label{eq:gain_mismatch}
\end{align}}
{{For $r\geq3$, \eqref{eq:gain_mismatch} is the scalar specialization of \cite[Sec.~3.2]{lanza2022output}. The midpoint $\widetilde\gamma=(\underline\gamma+\overline\gamma)/2$ is feasible whenever $(\overline\gamma-\underline\gamma)/(\overline\gamma+\underline\gamma)<\bar\mu_{r,\rho}$.}}

For $r\geq2$, write $\xi_i:=\col(\xi_{i,1},\ldots,\xi_{i,r})\in\R^r$ for $i=1,\ldots,r-1$ and set $\xi_{0,1}:=e_F$. {{Set $\varphi_i:=\varphi_1$ for $i=1$ and, when $r\geq3$, set $\varphi_i:=\varphi_p$ for $i=2,\ldots,r-1$.}}
For $i=1,\ldots,r-1$, define the normalized interstage error
\[
 q_i:=\varphi_i(\xi_{i-1,1}-\xi_{i,1}),
\]
and, on the domain $\abs{q_i}<1$, use
\begin{align}
 h_i&:=\frac{1}{1-q_i^2},\nonumber\\
 \dot\xi_{i,j}
 &=\xi_{i,j+1}+(a_j+p_jh_i)(\xi_{i-1,1}-\xi_{i,1}),
 &&j=1,\ldots,r-1,\label{eq:fpc1}\\
 \dot\xi_{i,r}
 &=\widetilde\gamma v+(a_r+p_rh_i)(\xi_{i-1,1}-\xi_{i,1}).\label{eq:fpc2}
\end{align}
Define the terminal artificial output for every $r\geq1$ by
\begin{equation}
 z:=\begin{cases}e_F,&r=1,\\\xi_{r-1,1},&r\geq2.\end{cases}
\end{equation}
For $r\geq2$, the derivatives $z,\dot z,\ldots,z^{(r-1)}$ are computed recursively from $e_F$, the FPC states, and the prescribed funnel functions, without differentiating $e_F$ and without an algebraic dependence on $v$ \cite{berger2018fpc,lanza2022output}.

For $\eta=(\eta_1,\ldots,\eta_k)\in\R^k$ and $1\leq j\leq k$, write $\eta_{1:j}:=(\eta_1,\ldots,\eta_j)$. Define
\[
 \mathcal D_1:=(-1,1),\qquad
 \sigma:\mathcal D_1\to\R,\qquad
 \sigma(q):=\frac{q}{1-q^2},
\]
and set $\mathcal R_1:\mathcal D_1\to\mathcal D_1$, $\mathcal R_1(\eta_1):=\eta_1$. For $k=2,\ldots,r$ and every $\eta\in\R^k$ satisfying $\eta_{1:k-1}\in\mathcal D_{k-1}$, define

\begin{align}
 \mathcal R_k(\eta)
 &:=\eta_k+\sigma\!\left(\mathcal R_{k-1}(\eta_{1:k-1})\right),\\
 \mathcal D_k
 &:=\bigl\{\eta\in\R^k:\ \eta_{1:k-1}\in\mathcal D_{k-1},\nonumber\\[-1mm]
 &\hspace{34mm}\abs{\mathcal R_k(\eta)}<1\bigr\}.
\end{align}

For $\phi\in\Phi_1$, set
\begin{equation}
 \eta^z:=(\phi z,\phi\dot z,\ldots,\phi z^{(r-1)}).
\end{equation}
The performance funnel associated with the terminal artificial output $z$ and the function $\phi$ is called the terminal funnel. It is open at the initial time when $\phi(0)=0$, in which case its reciprocal boundary is unbounded as $t\downarrow0$. Every $\phi\in\Phi_1$ satisfies the terminal-funnel regularity assumptions used in \cite[Sec.~4]{lanza2022output} and \cite[Thm.~1.9]{berger2021nonlinear}. On the domain $\eta^z\in\mathcal D_r$, define the terminal recursive coordinate and terminal gain by

\begin{align}
 \omega&:=\mathcal R_r(\eta^z),\nonumber\\
 h_c&:=\frac{1}{1-\omega^2},\qquad
 v:=-\sigma(\omega)=-h_c\omega.
 \label{eq:funnel_controller}
\end{align}

For $r=1$, this reduces to $\omega=\phi e_F$.

\begin{figure*}[t]
\centering
{
\resizebox{0.80\textwidth}{!}{%
\begin{tikzpicture}[
  font=\footnotesize,
  block/.style={draw=black,fill=white,rectangle,minimum height=8.5mm,minimum width=18mm,align=center,inner sep=2pt},
  resblock/.style={draw=black,fill=white,rectangle,minimum height=12mm,minimum width=26mm,align=center,inner sep=2.6pt},
  fpcblock/.style={draw=black,fill=white,rectangle,minimum height=11.5mm,minimum width=26mm,align=center,inner sep=2.6pt,font=\scriptsize},
  termblock/.style={draw=black,fill=white,rectangle,minimum height=14.5mm,minimum width=28mm,align=center,inner sep=2.6pt,font=\scriptsize},
  sum/.style={draw=black,circle,minimum size=6mm,inner sep=0pt},
  branch/.style={circle,fill=black,inner sep=1.05pt},
  arrow/.style={draw=black,-Latex,semithick},
  wire/.style={draw=black,semithick},
  lab/.style={font=\footnotesize,text=black,fill=white,inner sep=1pt},
  slabel/.style={font=\scriptsize,text=black,fill=white,inner sep=0.7pt}
]
\coordinate (w0) at (0,0);
\node[sum] (s1) at (1.15,0) {};
\node[block] (C) at (3.25,0) {$C$};
\coordinate (uctap) at (5.35,0); \node[branch] at (uctap) {};
\node[sum] (s2) at (7.25,0) {};
\node[sum] (s3) at (9.85,0) {};
\node[block,minimum width=23mm] (P) at (12.75,0) {$P_\theta$};
\coordinate (ytap) at (16.80,0); \node[branch] at (ytap) {};
\coordinate (yout) at (17.70,0);

\draw[arrow] (w0)--node[lab,above]{$w$}(s1.west);
\draw[arrow] (s1.east)--node[lab,above]{$w-y$}(C.west);
\draw[wire]  (C.east)--node[lab,above]{$u_c$}(uctap);
\draw[arrow] (uctap)--(s2.west);
\draw[arrow] (s2.east)--node[lab,above]{$u$}(s3.west);
\draw[arrow] (s3.east)--node[lab,above]{$u+d$}(P.west);
\draw[wire]  (P.east)--(ytap);
\draw[arrow] (ytap)--node[lab,above]{$y$}(yout);

\node[font=\scriptsize,text=black] at ($(s1.west)+(1.45mm,0)$) {$+$};
\node[font=\scriptsize,text=black] at ($(s1.south)+(0,1.45mm)$) {$-$};
\node[font=\scriptsize,text=black] at ($(s2.west)+(1.45mm,0)$) {$+$};
\node[font=\scriptsize,text=black] at ($(s2.south)+(0,1.45mm)$) {$+$};
\node[font=\scriptsize,text=black] at ($(s3.west)+(1.45mm,0)$) {$+$};
\node[font=\scriptsize,text=black] at ($(s3.north)+(0,-1.45mm)$) {$+$};

\coordinate (dtop) at ($(s3.north)+(0,4mm)$);
\draw[arrow] (dtop)--node[lab,right]{$d$}(s3.north);

\coordinate (boxNW) at (3.90,-1.1);
\coordinate (boxSE) at (16.0,-4.47);
\draw[black,dashed,rounded corners=1pt] (boxNW) rectangle (boxSE);
\node[anchor=north west,fill=white,inner sep=1pt,text=black]
  at ($(boxNW)+(1.70,-0.03)$) {Proposed FPC-DOB};

\node[resblock] (R) at (5.35,-2.72) {Residual filter\\[-0.4mm]$e_F=M_ny-N_nu_c$};
\node[fpcblock] (FPC) at (8.80,-2.72) {FPC cascade\\[-0.4mm]$\xi_1,\ldots,\xi_{r-1}$};
\node[termblock] (TF) at (13.55,-2.72) {Terminal funnel law\\[-0.4mm]$\omega=\mathcal R_r(\eta^z)$\\[-0.4mm]$v=-\sigma(\omega)$};
\coordinate (vtap) at (15.40,-2.72); \node[branch] at (vtap) {};

\draw[arrow] (uctap)--node[slabel,right,pos=.72]{$u_c$}(R.north);

\coordinate (ySplit) at ($(ytap)+(0,-4.08)$);
\node[branch] at (ySplit) {};
\draw[wire] (ytap)--(ySplit);
\draw[arrow] (ySplit) -| (R.south);

\draw[arrow] (R.east)--node[lab,above]{$e_F$}(FPC.west);
\draw[arrow] (FPC.east)--node[slabel,above,pos=.50]{$z,\dot z,\ldots,z^{(r-1)}$}(TF.west);

\draw[wire] (TF.east)--(vtap);
\node[lab,right] at ($(vtap)+(0.28,0)$) {$v$};

\coordinate (vExtLane) at ($(s2.south)+(0,-0.36)$);
\draw[arrow] (vtap) |- (vExtLane) -- (s2.south);

\coordinate (vIntLane) at ($(FPC.south)+(0,-0.38)$);
\draw[arrow] (vtap) |- (vIntLane) -- (FPC.south);

\coordinate (fbLane) at ($(s1.south)+(0,-4.38)$);
\draw[arrow] (ySplit) |- (fbLane) -- (s1.south);
\end{tikzpicture}}}
\caption{{Implemented FPC-DOB architecture. The standard outer feedback loop is retained, while the proposed FPC-DOB forms the residual $e_F=M_ny-N_nu_c$, passes it through the FPC cascade, and generates the compensation $v$ through the terminal funnel law. For $r=1$, $z=e_F$ and the FPC cascade is bypassed.}}
\label{fig:architecture}
\end{figure*}
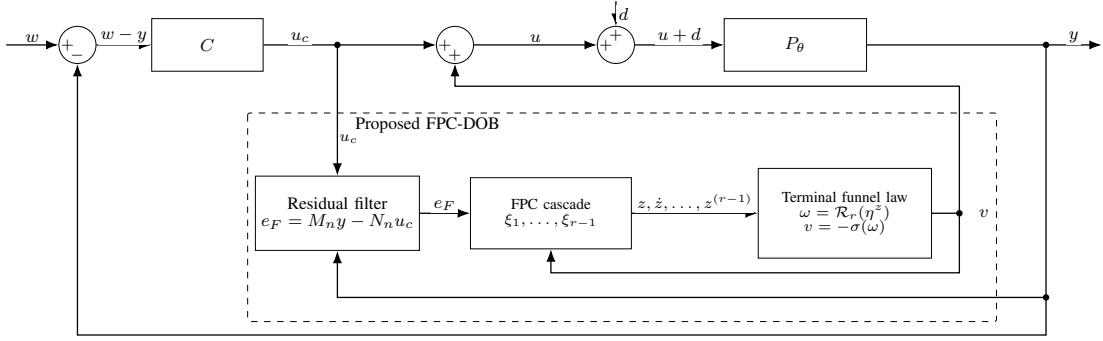

{Figure~\ref{fig:architecture} summarizes the implemented interconnection.}

The FPC states may be initialized arbitrarily subject to the following domain conditions. {{The applicable interstage conditions are}}
{
\begin{align}
 \varphi_1(0)\abs{e_F(0)-\xi_{1,1}(0)}&<1,
 &&r\geq2,\label{eq:init1}\\
 \varphi_p(0)\abs{\xi_{i-1,1}(0)-\xi_{i,1}(0)}&<1,
 &&i=2,\ldots,r-1,\quad r\geq3.\label{eq:init2}
\end{align}}
For every $r\geq1$, assume
\begin{equation}
 \eta^z(0)\in\mathcal D_r.
 \label{eq:init3}
\end{equation}

For $t>0$, define
{
\begin{equation}
 b_F(t):=
 \begin{cases}
  \phi(t)^{-1}, & r=1,\\[1mm]
  \displaystyle\sum_{i=1}^{r-1}\varphi_i(t)^{-1}+\phi(t)^{-1}, & r\geq2.
 \end{cases}
\end{equation}}

\begin{theorem}[Globality, boundedness, and residual confinement]\label{thm:main}
Consider the complete implemented closed loop consisting of \eqref{eq:plant}, \eqref{eq:nominal_controller}, \eqref{eq:input}, \eqref{eq:residual}, \eqref{eq:funnel_controller}, and, when $r\geq2$, \eqref{eq:fpc1}--\eqref{eq:fpc2}. Let $\phi\in\Phi_1$. {{For $r\geq2$, let $\varphi_1\in\Phi_r$, select $a_1,\ldots,a_r>0$ so that $A_f$ is Hurwitz, choose $Q_f\succ0$, solve \eqref{eq:Af} for $X_f$, define $p_1,\ldots,p_r$ through \eqref{eq:p}, and choose one common $\widetilde\gamma>0$. When $r\geq3$, additionally let $\varphi_p=\rho\varphi_1$ for $\rho>1$ and require \eqref{eq:gain_mismatch}.}} Suppose Assumptions~\ref{ass:plants}--\ref{ass:nominal_loop} hold and that $w,d\in\Linfty$. Let the plant, controller, and stable-factor states start at zero, and {{assume \eqref{eq:init3}, \eqref{eq:init1} when $r\geq2$, and \eqref{eq:init2} when $r\geq3$.}}

For each $\theta\in\Theta$, the unique maximal Carath\'eodory solution extends to $[0,\infty)$; all plant, controller, stable-factor, and FPC states, the signals $e_F,z,v,\hat d_F,y,u_c,u$, the terminal gain $h_c$, and, for $r\geq2$, the gains $h_1,\ldots,h_{r-1}$ are bounded. The same $\Lambda$ and terminal profile $\phi$ are used for every $\theta\in\Theta$. {{When $r\geq2$, the same $\varphi_1$, coefficients $a_j,p_j$, and common $\widetilde\gamma$ are used; when $r\geq3$, the same $\varphi_p$ and ratio $\rho$ are also used.}} Only the resulting trajectory bounds may depend on $\theta$.

If $r\geq2$, there exist $\varepsilon_1,\ldots,\varepsilon_{r-1},\beta>0$ such that, for all $t>0$,
{
\begin{align}
 \abs{\xi_{i-1,1}(t)-\xi_{i,1}(t)}
 &<\varphi_i(t)^{-1}-\varepsilon_i,
 &&i=1,\ldots,r-1,\nonumber\\
 \abs{z(t)}&<\phi(t)^{-1}-\beta.
 \label{eq:individual_bounds}
\end{align}}

If $r\geq2$, set $\varepsilon_F:=\sum_{i=1}^{r-1}\varepsilon_i+\beta$. If $r=1$, set
\[
 \kappa_F:=\norm{\phi e_F}_{\infty}<1,
 \qquad
 \varepsilon_F:=\frac{1-\kappa_F}{2\norm{\phi}_{\infty}}>0.
\]
In either case, for all $t>0$,
\begin{equation}
 \abs{e_F(t)}<b_F(t)-\varepsilon_F.
 \label{eq:residual_bound}
\end{equation}

\end{theorem}

\begin{proof}
For $r\geq2$, Lemma~\ref{lem:channel} gives a linear minimum-phase residual channel of relative degree $r$, high-frequency gain $\gamma_\theta$, and bounded matched additive input. By \cite[Rem.~3.7]{lanza2022output}, the minimal realization \eqref{eq:residual_realization} of $G_\theta$ belongs to the linear subclass of the FPC system class. {{For $r=2$, \cite[Thm.~2]{berger2018fpc} applies under $\gamma_\theta,\widetilde\gamma>0$ with no mismatch condition; neither $\varphi_p$ nor $\rho$ is required. For $r\geq3$, the Lyapunov construction \eqref{eq:Af}--\eqref{eq:p}, the relation $\varphi_p=\rho\varphi_1$, and \eqref{eq:gain_mismatch} satisfy the scalar conditions in \cite[Sec.~3.2]{lanza2022output}. In both cases, the corresponding preservation result shows that the augmented FPC channel remains minimum phase with relative degree $r$.}} With the preservation result established above, we follow the proof of \cite[Cor.~4.2]{lanza2022output} for zero reference. The initialization $x_R(0)=0$ and \eqref{eq:markov_parameters} give
\[
 e_F^{(k)}(0)=C_RA_R^kx_R(0)=0,\qquad k=0,\ldots,r-1.
\]
{{With the constant zero residual-output history, this jet, \eqref{eq:init1}, \eqref{eq:init3}, and, when $r\geq3$, \eqref{eq:init2} provide admissible data.}} Hence every maximal solution of the residual-channel/FPC/terminal-controller interconnection is global. The same argument also gives boundedness of $v$, the FPC states, the interstage gains $h_i$, and $e_F,\dot e_F,\ldots,e_F^{(r-1)}$, together with the margins in \eqref{eq:individual_bounds}. Since $v=-\omega/(1-\omega^2)$ is bounded while $\abs{\omega}<1$, $\omega$ is uniformly separated from $\{-1,1\}$ and $h_c$ is bounded. The telescoping identity
\[
 e_F=(e_F-\xi_{1,1})+
 \sum_{i=2}^{r-1}(\xi_{i-1,1}-\xi_{i,1})+z
\]
then gives \eqref{eq:residual_bound}.

For $r=1$, the realization \eqref{eq:residual_realization} is a SISO linear minimum-phase system of relative degree one with bounded matched additive input and positive high-frequency gain {$\gamma_\theta$}; hence it belongs to the known-sign subclass of \cite[Thm.~1.9]{berger2021nonlinear}. Applied to \eqref{eq:residual_realization} with $x_R(0)=0$, that result yields a global residual-channel/controller solution, bounded $v$ and $e_F$, and $\kappa_F=\norm{\phi e_F}_{\infty}<1$. Consequently $h_c$ is bounded. Moreover, for every $t>0$,
\[
 \phi(t)^{-1}-\abs{e_F(t)}
 =\frac{1-\abs{\phi(t)e_F(t)}}{\phi(t)}
 \geq\frac{1-\kappa_F}{\norm{\phi}_{\infty}}
 =2\varepsilon_F,
\]
which proves \eqref{eq:residual_bound}.

The interconnection of \eqref{eq:residual_realization}, the applicable FPC cascade, and the terminal controller therefore has a global solution with the prescribed FPC initial states. On the open domain determined by $\eta^z\in\mathcal D_r$ and, when $r\geq2$, $\abs{q_i}<1$, the complete implemented closed-loop right-hand side is measurable in time and locally Lipschitz in the full state; hence local Carath\'eodory uniqueness holds.

Driven by the global compensation signal $v$ generated by that interconnection and by $w,d$, the physical plant--controller--stable-factor subsystem is linear and has a unique solution on every finite interval. Because its plant, controller, and stable-factor states are initialized at zero, its residual equals $G_\theta(v+\bar d_\theta)$. Coupling this physical trajectory with the corresponding global FPC states gives a global solution of the complete implemented closed loop, and local uniqueness identifies it with the unique maximal physical solution.

Using \eqref{eq:npr_map}, the transfer functions
\[
 T_n:=\frac{P_nC}{1+P_nC}=\frac{B_nN_c}{\chi_n},\qquad
 H_y=\frac{\Lambda D_c}{\chi_n},
\]
\[
 CH_y=\frac{N_c\Lambda}{\chi_n},\qquad
 \frac{C}{1+P_nC}=\frac{A_nN_c}{\chi_n}
\]
are stable and proper. Thus $y_0=T_nw$ is bounded, $y=y_0+H_ye_F$ is bounded, and
\[
 u_c=\frac{C}{1+P_nC}w-CH_ye_F
\]
is bounded. The input $u=u_c+v$ is consequently bounded. To bound the actual plant and nominal-controller states, consider a minimal SISO LTI realization with state dimension $n_g$,
\[
 \dot x=A_g x+B_g u_g,\qquad
 y_g=C_g x+D_g u_g,
\]
where $A_g\in\R^{n_g\times n_g}$, $B_g\in\R^{n_g}$, $C_g\in\R^{1\times n_g}$, and $D_g\in\R$.
Observability gives $T_o>0$ such that
\[
 W_o(T_o):=\int_0^{T_o}
 e^{A_g^\top \nu}C_g^\top C_g e^{A_g\nu}\,\dd\nu
 \succ0.
\]
For $t\geq T_o$ and $\nu\in[0,T_o]$, define
\[
\begin{aligned}
 \widetilde y_t(\nu):={}&y_g(t-T_o+\nu)-D_gu_g(t-T_o+\nu)\\
 &-\int_0^\nu C_ge^{A_g(\nu-\ell)}B_gu_g(t-T_o+\ell)\,\dd\ell.
\end{aligned}
\]
The variation-of-constants formula gives
\[
 \widetilde y_t(\nu)=C_ge^{A_g\nu}x(t-T_o),
\]
and therefore
\[
 x(t-T_o)=W_o(T_o)^{-1}
 \int_0^{T_o}e^{A_g^\top\nu}C_g^\top\widetilde y_t(\nu)\,\dd\nu.
\]
Propagating this reconstructed state through
\[
 x(t)=e^{A_gT_o}x(t-T_o)
 +\int_0^{T_o}e^{A_g(T_o-\nu)}B_gu_g(t-T_o+\nu)\,\dd\nu
\]
gives constants $c_y,c_u>0$ such that, for every $t\geq T_o$,
\[
 \norm{x(t)}\leq
 c_y\operatorname*{ess\,sup}_{\nu\in[t-T_o,t]}\abs{y_g(\nu)}
 +c_u\operatorname*{ess\,sup}_{\nu\in[t-T_o,t]}\abs{u_g(\nu)}.
\]
Apply this estimate to the actual input--output pair $(u_g,y_g)=(u+d,y)$ of the plant and, when the nominal controller has positive state dimension, to the pair $(u_g,y_g)=(w-y,u_c)$ of its minimal realization. A static controller has no realization state to bound. The $M_n$ and $N_n$ realizations have Hurwitz state matrices determined by $\Lambda$ and bounded inputs $y$ and $u_c$, so their states are bounded by variation of constants. The same variation-of-constants formula directly bounds the plant and controller states on the initial interval $[0,T_o]$; for $t\geq T_o$, the finite-horizon estimate gives uniform bounds. Thus the plant, controller, and stable-factor states are uniformly bounded on $[0,\infty)$, completing the state-boundedness argument.
\end{proof}

\begin{corollary}[Nominal-output recovery and weighted-estimation bounds]\label{cor:recovery_bounds}
Under the assumptions of Theorem~\ref{thm:main}, additionally suppose $b_F\in\Linfty$. Write the impulse response of the stable recovery map $H_y$ in \eqref{eq:npr_map} as $h_y=h_0\delta_{\mathrm D}+\bar h_y$, where $h_0\in\R$ and $\bar h_y\in\Lone$. {Then, for every $t>0$, the deviation from the nominal output $y_0$ defined in \eqref{eq:y0} satisfies}
\begin{equation}
 \abs{y(t)-y_0(t)}
 \leq\abs{h_0}b_F(t)
 +\int_0^t\abs{\bar h_y(t-\ell)}b_F(\ell)\,\dd\ell.
 \label{eq:npr_pointwise}
\end{equation}
Moreover,
\begin{equation}
 \norm{y-y_0}_{\infty}
 \leq\norm{H_y}_{\Lone}\norm{b_F}_{\infty}.
 \label{eq:npr_bound}
\end{equation}
Under the same condition, choose any $W\in\RH$ satisfying $WN_n^{-1}\in\RH$ and use zero initial states for both stable weighting filters. Define the weighted lumped-disturbance signal $d_{W,\theta}:=WN_n^{-1}\zeta_F$, its estimate $\hat d_W:=W\hat d_F$, and the corresponding error $e_{d,W}:=d_{W,\theta}-\hat d_W=WN_n^{-1}e_F$. Then
\begin{equation}
 \norm{e_{d,W}}_{\infty}
 \leq\norm{WN_n^{-1}}_{\Lone}\norm{b_F}_{\infty}.
 \label{eq:weighted_estimation}
\end{equation}
\end{corollary}

\begin{proof}
Since $h_y=h_0\delta_{\mathrm D}+\bar h_y$,
\[
 y(t)-y_0(t)
 =h_0e_F(t)+\int_0^t\bar h_y(t-\ell)e_F(\ell)\,\dd\ell.
\]
Using $\abs{e_F(\ell)}<b_F(\ell)$ gives \eqref{eq:npr_pointwise}. Taking the essential supremum and applying the $\Lone$-$\Linfty$ convolution inequality yields
\[
 \norm{y-y_0}_{\infty}
 \leq\bigl(\abs{h_0}+\norm{\bar h_y}_{\Lone}\bigr)
 \norm{b_F}_{\infty}
 =\norm{H_y}_{\Lone}\norm{b_F}_{\infty}.
\]
Applying the stable convolution operator $WN_n^{-1}$ to $e_F$ gives \eqref{eq:weighted_estimation}.
\end{proof}

\begin{remark}[Weighted-error interpretation and nominal transparency]
For every admissible $W$ in Corollary~\ref{cor:recovery_bounds}, if $P_\theta=P_n$, then $\zeta_F=N_nd$, so $d_{W,\theta}=Wd$ and $e_{d,W}=W(d-\hat d_F)$. In particular, because $P_n$ is minimum phase and has relative degree $r$, $W_{\lambda_W}(s):=(\lambda_W/(s+\lambda_W))^r$, $\lambda_W>0$, satisfies $W_{\lambda_W}N_n^{-1}\in\RH$.

A separate consequence is nominal transparency: when $P_\theta=P_n$, $d=0$, and the plant, nominal-controller, stable-factor, and all applicable FPC states are initialized at zero, the FPC-DOB augmentation preserves the nominal closed-loop trajectory. Consider the candidate augmented trajectory formed by that nominal plant--controller response, zero stable-factor initial states, and FPC states that remain identically zero. Along this trajectory, \eqref{eq:residual} gives $e_F\equiv0$. Hence every interstage error is zero when $r\geq2$, so $q_i=0$, $h_i=1$, and the FPC state derivatives vanish recursively. Also, $z=\dot z=\cdots=z^{(r-1)}=0$, hence $\eta^z=0\in\mathcal D_r$, $\omega=0$, $h_c=1$, and $v=\hat d_F=0$, independently of the time variation of the funnel functions. The candidate satisfies every closed-loop equation, and local Carath\'eodory uniqueness preserves it; consequently $u=u_c$ identically.
\end{remark}

\begin{corollary}[A posteriori gain and compensation certificate]\label{cor:gain_certificate}
For every closed-loop trajectory satisfying the assumptions of Theorem~\ref{thm:main}, define $\kappa_c:=\norm{\omega}_{\infty}$. Then
\begin{equation}
 \kappa_c<1,\qquad
 {\norm{h_c}_{\infty}=\frac{1}{1-\kappa_c^2}},\qquad
 {\norm{v}_{\infty}=\frac{\kappa_c}{1-\kappa_c^2}}.
\end{equation}
When $r\geq2$, additionally define
\[
 \kappa_i:=\norm{q_i}_{\infty},
 \quad i=1,\ldots,r-1.
\]
Then $\kappa_i<1$ and
\begin{equation}
 {\norm{h_i}_{\infty}=\frac{1}{1-\kappa_i^2}},
 \qquad i=1,\ldots,r-1.
\end{equation}
\end{corollary}

\begin{proof}
{For $q=\omega$, let $g_q:=h_c$, and, when $r\geq2$, for $q=q_i$, let $g_q:=h_i$. Boundedness of $g_q=(1-q^2)^{-1}$ together with $\abs{q(t)}<1$ a.e. gives $\kappa:=\norm{q}_\infty<1$. Since $s\mapsto(1-s^2)^{-1}$ and $s\mapsto s(1-s^2)^{-1}$ are continuous and increasing on $[0,1)$, taking essential suprema gives the stated equalities.}
\end{proof}

\begin{remark}[Finite initial boundaries]
For the continuous funnel classes used here, $b_F\in\Linfty$ holds precisely when every participating funnel function is positive at $t=0$. Finite initial boundaries therefore yield \eqref{eq:npr_pointwise}--\eqref{eq:weighted_estimation}. For open initial funnels, $b_F\notin\Linfty$, while Theorem~\ref{thm:main} still gives globality, boundedness, and residual confinement.
\end{remark}

For $R_0,R_\infty,\lambda>0$, define
\begin{equation}
 \psi_{R_0,R_\infty,\lambda}(t)
 :=\frac{1-e^{-\lambda t}}{R_\infty+R_0e^{-\lambda t}}.
 \label{eq:open_profile}
\end{equation}
Writing $\psi:=\psi_{R_0,R_\infty,\lambda}$, this profile belongs to $\Phi_k$ for every integer $k\geq1$, satisfies $\psi(0)=0$, and obeys
\[
\begin{aligned}
 \psi(t)^{-1}
 &=\frac{R_\infty+R_0}{\lambda t}+O(1),
 &&t\downarrow0,\\
 \psi(t)^{-1}&\longrightarrow R_\infty,
 &&t\to\infty.
\end{aligned}
\]
Thus $R_\infty+R_0$ and $\lambda$ determine the leading near-origin growth of the funnel boundary, whereas $R_\infty$ is its asymptotic radius. {{Profiles of the form \eqref{eq:open_profile} may be chosen separately for $\varphi_1$ and $\phi$. When $r\geq3$, one additionally sets $\varphi_p=\rho\varphi_1$.}}

{The zero-state convention applies only to transfer-function identities for the implemented plant, controller, and stable-factor realizations. Open-funnel initialization instead allows arbitrary finite residual-channel and FPC initial states. Finite initial boundaries require the compatibility conditions \eqref{eq:init1}--\eqref{eq:init3}; open initial profiles make every normalized initial coordinate zero, so those conditions hold automatically and the controller starts from unit gains and zero correction.}

\begin{proposition}[Open-funnel initialization without compatibility restrictions]\label{prop:open_initialization}
Fix $\theta\in\Theta$ and consider the residual-channel/FPC/terminal-controller interconnection consisting of \eqref{eq:residual_realization}, \eqref{eq:fpc1}--\eqref{eq:fpc2} when $r\geq2$, and \eqref{eq:funnel_controller}. {{Use $\phi\in\Phi_1$ and, when $r\geq2$, the applicable interstage profiles $\varphi_i\in\Phi_r$ defined above.}} Suppose Assumptions~\ref{ass:plants}--\ref{ass:nominal_loop} hold, $w,d\in\Linfty$, $x_R(0)\in\R^{n_R}$ is the initial state of \eqref{eq:residual_realization}, and all applicable FPC initial states are arbitrary vectors in their Euclidean state spaces. {{Let $\varphi_i(0)=0$, $i=1,\ldots,r-1$, when $r\geq2$, and let $\phi(0)=0$.}} Then all applicable initial-domain conditions \eqref{eq:init1}--\eqref{eq:init3} hold automatically. There exist $T>0$ and positive constants $c_c,c_v$ such that, for $0\leq t\leq T$,
\begin{equation}
 0\leq h_c(t)-1\leq c_ct^2,\qquad \abs{v(t)}\leq c_vt.
 \label{eq:local_v}
\end{equation}
When $r\geq2$, there also exist $c_i>0$, $i=1,\ldots,r-1$, such that
\begin{equation}
 0\leq h_i(t)-1\leq c_it^2,\qquad i=1,\ldots,r-1.
 \label{eq:local_hi}
\end{equation}
In particular, $h_c(0)=1$ and $v(0)=0$; when $r\geq2$, $h_i(0)=1$, $i=1,\ldots,r-1$. The constants may depend on the fixed plant, the fixed FPC design coefficients, the norms of the selected initial states, local essential bounds of $w,d$, and the selected funnel profiles.
\end{proposition}

\begin{proof}
Since $\varphi_i(0)=0$ when $r\geq2$, every normalized interstage error satisfies $q_i(0)=0$, and hence $h_i(0)=1$. Equation~\eqref{eq:markov_parameters} gives the finite initial jet $e_F^{(k)}(0)=C_RA_R^kx_R(0)$, $k=0,\ldots,r-1$, without derivatives of $w$ or $d$; \eqref{eq:fpc1}--\eqref{eq:fpc2} then recursively express $z(0),\dot z(0),\ldots,z^{(r-1)}(0)$ as finite algebraic functions of the initial states, design coefficients, and prescribed profiles. For $r=1$, $z(0)=e_F(0)=C_Rx_R(0)$. In either case, $\phi(0)=0$ gives $\eta^z(0)=0\in\mathcal D_r$. The recursion for $\mathcal R_k$ then gives $\omega(0)=0$, $h_c(0)=1$, and $v(0)=0$.

Let $x_{\mathrm{aug}}:=\col(x_R,\xi_1,\ldots,\xi_{r-1})$ when $r\geq2$ and $x_{\mathrm{aug}}:=x_R$ when $r=1$. On the open domain defined by $\eta^z\in\mathcal D_r$ and, when $r\geq2$, $\abs{q_i}<1$, $i=1,\ldots,r-1$, the right-hand side of the finite-dimensional interconnection is measurable in time and locally Lipschitz in $x_{\mathrm{aug}}$. The standard Carath\'eodory existence and uniqueness theorem therefore gives a unique local solution {\cite{walter1998}}. Along a sufficiently short interval $[0,T]$, the residual-channel and FPC states remain bounded. Equations~\eqref{eq:fpc1}--\eqref{eq:fpc2} then make $z^{(j)}$, $j=0,\ldots,r-1$, locally bounded when $r\geq2$; for $r=1$, $z=e_F$ is locally bounded.

Let
\[
 L_\phi:=\operatorname*{ess\,sup}_{\tau\in[0,T]}\abs{\dot\phi(\tau)}<\infty.
\]
Since $\phi(0)=0$,
\[
 \abs{\phi(t)}
 \leq\int_0^t\abs{\dot\phi(\tau)}\,\dd\tau
 \leq L_\phi t.
\]
Thus $\phi(t)=O(t)$, and the same argument gives $\varphi_i(t)=O(t)$ for the open FPC profiles. Choose $M_j>0$ such that $\abs{z^{(j)}(t)}\leq M_j$ on $[0,T]$ for $j=0,\ldots,r-1$. Then
\[
 \abs{\phi(t)z^{(j)}(t)}\leq M_jL_\phi t,
\]
so every component of $\eta^z(t)$ is $O(t)$. Local boundedness of the interstage errors similarly gives $q_i(t)=O(t)$.

Next, $\mathcal R_1(\eta_1^z)=\eta_1^z=O(t)$. If $\mathcal R_{k-1}(\eta_{1:k-1}^z)=O(t)$, then $\sigma(0)=0$ and local Lipschitz continuity of $\sigma$ at zero give
\[
 \sigma\!\left(\mathcal R_{k-1}(\eta_{1:k-1}^z)\right)=O(t).
\]
Since $\eta_k^z=O(t)$, the recursion gives $\mathcal R_k(\eta_{1:k}^z)=O(t)$. Hence $\omega(t)=O(t)$. Choose $L_c>0$ and reduce $T$ so that $\abs{\omega(t)}\leq L_ct$ and $\abs{\omega(t)}\leq1/2$ on $[0,T]$. When $r\geq2$, choose $L_i>0$ and reduce $T$ further so that $\abs{q_i(t)}\leq L_it$ and $\abs{q_i(t)}\leq1/2$. Then
\[
 \begin{aligned}
 h_c(t)-1
 &=\frac{\omega(t)^2}{1-\omega(t)^2}
 \leq\frac{4}{3}L_c^2t^2,\\
 \abs{v(t)}
 &=\frac{\abs{\omega(t)}}{1-\omega(t)^2}
 \leq\frac{4}{3}L_ct.
 \end{aligned}
\]
For $r\geq2$,
\[
 h_i(t)-1
 =\frac{q_i(t)^2}{1-q_i(t)^2}
 \leq\frac{4}{3}L_i^2t^2.
\]
Thus \eqref{eq:local_v} holds with $c_c=4L_c^2/3$ and $c_v=4L_c/3$. When $r\geq2$, \eqref{eq:local_hi} holds with $c_i=4L_i^2/3$, $i=1,\ldots,r-1$.
\end{proof}

\begin{corollary}[Open-funnel practical recovery]\label{cor:open_recovery}
For the FPC-DOB of Theorem~\ref{thm:main}, fix $\theta\in\Theta$, $\delta>0$, and any applicable FPC initial states. Using the profile \eqref{eq:open_profile}, choose positive $r_{c0},\bar r_c,\lambda_c$ and set $\phi(t):=\psi_{r_{c0},\,\delta\bar r_c,\,\lambda_c}(t)$. When $r\geq2$, additionally choose positive $r_{10},\bar r_1,\lambda_1$ and set $\varphi_1(t):=\psi_{r_{10},\,\delta\bar r_1,\,\lambda_1}(t)$. {{When $r\geq3$, set $\varphi_p(t):=\rho\varphi_1(t)$.}} Define
{
\[
 \bar B_r:=\begin{cases}
 \bar r_c,&r=1,\\
 \bar r_1+\bar r_c,&r=2,\\
 (1+(r-2)/\rho)\bar r_1+\bar r_c,&r\geq3.
 \end{cases}
\]}
{{Assume all conditions of Theorem~\ref{thm:main} except the applicable initial-domain conditions \eqref{eq:init1}--\eqref{eq:init3}.}} Proposition~\ref{prop:open_initialization} verifies the initial domains. The residual bound of Theorem~\ref{thm:main}, combined with the stable recovery map, gives
\begin{equation}
 \limsup_{t\to\infty}\abs{y(t)-y_0(t)}
 \leq\delta\norm{H_y}_{\Lone}\bar B_r.
\end{equation}
\end{corollary}

\begin{proof}
The selected profiles are open initially and satisfy $\limsup_{t\to\infty}b_F(t)\leq\delta\bar B_r$. Hence \eqref{eq:residual_bound} gives
\[
 \limsup_{t\to\infty}\abs{e_F(t)}\leq\delta\bar B_r.
\]
For the asymptotic output estimate, write the impulse response of $H_y$ as $h_y=h_0\delta_{\mathrm D}+\bar h_y$, where $\bar h_y\in\Lone$, and let $*$ denote causal convolution. Let
\[
 L:=\limsup_{t\to\infty}\abs{e_F(t)},\qquad
 M:=\norm{e_F}_{\infty}.
\]
For every $S>0$ and $t\geq S$,
\[
\begin{aligned}
 \abs{(\bar h_y*e_F)(t)}
 \leq{}&\int_0^S\abs{\bar h_y(\tau)}
       \abs{e_F(t-\tau)}\,\dd\tau\\
 &+M\int_S^\infty\abs{\bar h_y(\tau)}\,\dd\tau.
\end{aligned}
\]
Taking $\limsup_{t\to\infty}$ and then letting $S\to\infty$ gives
\[
 \limsup_{t\to\infty}\abs{(\bar h_y*e_F)(t)}
 \leq\norm{\bar h_y}_{\Lone}L.
\]
Including the direct term $h_0e_F$ in \eqref{eq:npr_map} yields
\[
 \limsup_{t\to\infty}\abs{y(t)-y_0(t)}
 \leq\bigl(\abs{h_0}+\norm{\bar h_y}_{\Lone}\bigr)L
 =\norm{H_y}_{\Lone}L.
\]
Substitution of $L\leq\delta\bar B_r$ proves the claim.
\end{proof}

\section{Numerical Evaluation}
{We examine initialization, disturbance rejection, and robustness to plant variation.} Consider
\begin{align}
 P_n&=\frac{1}{s^2+3s+2},&
 P_{\mathrm a}&=\frac{1}{s^2-3s+2},\nonumber\\
 C&=\frac{8(s+1)}{s+10}.&&
\end{align}
The polynomial $\chi_{\mathrm a}:=(s^2-3s+2)(s+10)+8(s+1)$ has two open-right-half-plane roots, whereas $\chi_n$ is Hurwitz; both plants are minimum phase, have relative degree two, and have high-frequency gain one. Use $P_{\mathrm a}$ as the actual plant, $P_n$ as the nominal model, $\Lambda=(s+1.5)^2$, $(a_1,a_2)=(6,9)$, $Q_f=I_2$, and $\mathbf p=[1,27/23]^\top$. {{For these second-order tests, $r=2$, so neither plant-gain bounds nor a mismatch condition is required; we set the common positive design gain $\widetilde\gamma=1$.}} For a uniform grid $\mathcal T$, write $\max_{\mathcal T}|g|:=\max_{t\in\mathcal T}|g(t)|$ and let $\operatorname{trapz}_{\mathcal T}$ denote composite trapezoidal integration.

{To compare finite and open funnels from the same initial state, set $w=d=0$ and all states to zero except $(\xi_{1,1}(0),\xi_{1,2}(0))=(0.2,1.4)$. Use the finite profiles
\[
 (\varphi_1^{\rm f})^{-1}=0.08+0.8e^{-0.7t},\qquad
 (\phi^{\rm f})^{-1}=0.07+0.6e^{-0.7t},
\]
and their open counterparts $\varphi_1^{\rm o}=\psi_{0.8,0.08,0.7}$ and $\phi^{\rm o}=\psi_{0.6,0.07,0.7}$. On $\mathcal T_p:=\{0,0.001,\ldots,3\}$, set $u_{{\rm corr},F}:=v$ and $u_{{\rm corr},Q}:=-\hat d_Q$. For each initialization run, let $u_{\rm corr}$ and $y$ denote the correction and output of its architecture, and define $U_0:=|u_{\rm corr}(0)|$, $U_p:=\max_{\mathcal T_p}|u_{\rm corr}|$, $t_p:=\min\operatorname*{arg\,max}_{t\in\mathcal T_p}|u_{\rm corr}(t)|$, and $Y_p:=\max_{\mathcal T_p}|y|$.}

{We next compare correction peaks of the open-profile FPC-DOB and a fixed-$Q$ DOB at matched sampled correction $L_2$ norm. Both use $w=d=0$; the fixed-$Q$ DOB uses $Q_\tau(s)=(\tau s+1)^{-2}$ at $\tau=0.10$, below the closed-loop stability boundary $\tau\approx0.117$ for $P_{\mathrm a}$. Let $x_F^0:=\col(\xi_{1,1}(0),\xi_{1,2}(0))$, let $x_Q^0$ be the initial state of a two-state controllable-canonical realization of the $Q_{0.10}u$ branch, and set all other states to zero. For $j\in\{F,Q\}$, write $u_{{\rm corr},j}(t;x_j^0)$ for the correction generated from $x_j^0$. To parameterize the initial-state directions, define
\[
\begin{aligned}
 K_j(t)&:=\left.\frac{\partial u_{{\rm corr},j}(t;x_j^0)}{\partial x_j^0}\right|_{x_j^0=0},\\
 \widehat{\mathcal G}_j&:=\operatorname{trapz}_{\mathcal T_p}(K_j^\top K_j).
\end{aligned}
\]
The computed matrices {{$\widehat{\mathcal G}_j$}} are positive definite. Set $E_*:=0.25$. For each angle $\vartheta\in[0,\pi)$, define $\mathbf e_\vartheta:=(\cos\vartheta,\sin\vartheta)^\top$ and, for a positive scale $\upsilon_j(\vartheta)$, $x_j^0(\vartheta):=\upsilon_j(\vartheta)\widehat{\mathcal G}_j^{-1/2}\mathbf e_\vartheta$. Write $u_j^\vartheta:=u_{{\rm corr},j}(\cdot;x_j^0(\vartheta))$ and select $\upsilon_j(\vartheta)$ to match the sampled correction norm $\sqrt{\operatorname{trapz}_{\mathcal T_p}((u_j^\vartheta)^2)}$ to $E_*$. The FPC-DOB absolute norm mismatch is at most $2\times10^{-5}$; {{by linearity and the normalization induced by $\widehat{\mathcal G}_Q$,}} set $\upsilon_Q(\vartheta):=E_*$ for the fixed-$Q$ DOB. The searches locally refine peak candidates from 24 FPC-DOB and 96 fixed-$Q$ angles; doubling these grids leaves the displayed maxima unchanged. Table~\ref{tab:initialization} summarizes both initialization comparisons.}

\begin{table}[H]
\caption{Sampled initialization metrics on $\mathcal T_p$}
\label{tab:initialization}
\centering
\scriptsize
\renewcommand{\arraystretch}{0.9}
\setlength{\tabcolsep}{1.5pt}
\begin{tabular}{@{}lrrrr@{}}
\hline
case & $U_0$ & $U_p$ & $t_p$ [s] & $Y_p$\\
\hline
finite FPC & 0.3449 & 0.5130 & 2.063 & 0.04756\\
open FPC & 0 & 0.3912 & 2.298 & 0.03704\\
open FPC, norm-matched & 0 & 0.3495 & 2.427 & 0.03294\\
$Q_{0.10}$-DOB, norm-matched & 0.9231 & 0.9231 & 0 & 0.00873\\
\hline
\end{tabular}
\end{table}
{The open profile gives $U_0=0$. At matched norm, the largest detected FPC-DOB peak is $0.3495$ versus $0.9231$ for the fixed-$Q$ DOB, about $62\%$ lower. The fixed-$Q$ DOB has smaller $Y_p$, so the result supports reduced correction peaking rather than uniform dominance.}

{For finite-profile disturbance rejection, use the following signals, where }$\mathbf1_I$ denotes the indicator function of any interval $I\subseteq\R_{\geq0}${:
\[
\begin{aligned}
 w_m&=1-e^{-0.8t}+0.15\sin(0.5t),\\
 d_m&=0.8\mathbf1_{[3,\infty)}+0.2\sin(1.5t).
\end{aligned}
\]
Use $(w,d)=(w_m,d_m)$ and $(\varphi_1,\phi)=(\varphi_1^{\rm f},\phi^{\rm f})$, initialize all states at zero, and set $b_F(0):=\lim_{t\downarrow0}b_F(t)$. States remain continuous across the jump at $t=3$. On $\mathcal T_m:=\{0,0.005,\ldots,20\}$, define
\[
\begin{aligned}
 \mathcal U_m&:=\max_{t\in\mathcal T_m}\frac{|e_F(t)|}{b_F(t)},\quad
 \mathcal M_m:=\min_{t\in\mathcal T_m}(b_F(t)-|e_F(t)|),\\
 \kappa_{1,m}&:=\max_{\mathcal T_m}|q_1|,\quad
 \kappa_{c,m}:=\max_{\mathcal T_m}|\omega|.
\end{aligned}
\]
The sampled values are
\[
 (\mathcal U_m,\mathcal M_m,\kappa_{1,m},\kappa_{c,m})
 =(0.6563,0.0516,0.8300,0.8225).
\]
These values show strict sampled margins, with nominal-output error $\max_{\mathcal T_m}|y-y_0|=0.2128$. Without compensation ($v\equiv0$), the same inputs and zero initialization give $|y(10)|=7.56\times10^4$.}

For a fixed-design sweep, define
\begin{equation}
 P_{\alpha}(s)=\frac{1}{s^2-\alpha s+2},\qquad
 \alpha\in\{-3\}\cup\{0,1,\ldots,6\}.
\end{equation}
{Here $P_n=P_{-3}$ and $P_{\mathrm a}=P_3$.} {Retain $w_m,d_m$, the finite profiles, all proposed parameters, zero initial states, and the fixed $Q_{0.10}$-DOB. On $\mathcal T_s:=\{0,0.01,\ldots,12\}$, let $y_F$ and $y_Q$ denote the proposed and fixed-$Q$ outputs, define $E_\alpha^F:=\max_{\mathcal T_s}|y_F-y_0|$ and $E_\alpha^Q:=\max_{\mathcal T_s}|y_Q-y_0|$, and let $\operatorname{sa}_Q(\alpha,\tau)$ denote the largest real part of the $Q_\tau$-DOB closed-loop state-matrix eigenvalues. Table~\ref{tab:sweep} reports $\alpha=0,\ldots,6$.}
\begin{table}[H]
\caption{Fixed-design uncertainty sweep on $\mathcal T_s$}
\label{tab:sweep}
\centering
\scriptsize
\renewcommand{\arraystretch}{0.9}
\setlength{\tabcolsep}{0.75pt}
\begin{tabular}{@{}lrrrrrrr@{}}
\hline
$\alpha$ & 0 & 1 & 2 & 3 & 4 & 5 & 6\\
\hline
$E_\alpha^F$ & 0.101 & 0.136 & 0.173 & 0.213 & 0.258 & 0.310 & 0.373\\
$E_\alpha^Q$ & 0.029 & 0.039 & 0.052 & 0.068 & 0.573 & $6.49{\times}10^2$ & $7.99{\times}10^5$\\
$\operatorname{sa}_Q(\alpha,0.10)$ & -0.794 & -0.756 & -0.725 & -0.411 & 0.212 & 0.818 & 1.411\\
\hline
\end{tabular}
\end{table}
{{At relative degree two, one common positive design gain may be selected without plant-gain bounds or a mismatch condition. In this no-retuning denominator-uncertainty sweep, the FPC-DOB trajectories remain bounded on $\mathcal T_s$, whereas the fixed $Q_{0.10}$-DOB is unstable for the tested $\alpha=4,5,6$. The stability boundaries of the same $Q_\tau$ family are approximately $\tau=0.093$ at $\alpha=4$ and $0.0664$ at $\alpha=6$, with the smaller-$\tau$ side stable.}}

Finally, to illustrate the construction at relative degrees $r=2,3,4$, set $P_{n,r}=(s+1)^{-r}$ and choose $C_r=N_{c,r}/D_{c,r}$ with $D_{c,r}$ monic of degree $r-1$, $\deg N_{c,r}\leq r-1$, $(D_{c,r},N_{c,r})$ coprime, and $(s+1)^rD_{c,r}+N_{c,r}=(s+2)^{2r-1}$. Set $(\mu_2,\mu_3,\mu_4)=(-0.4,-0.5,-0.6)$ and use $P_{{\rm a},r}=[(s+1)^r+(\mu_r-r)s^{r-1}]^{-1}$, $\Lambda_r=(s+2)^r$, $a_j=\binom{r}{j}3^j$, $j=1,\ldots,r$, $Q_f=I_r$, and $\widetilde\gamma=1$. {{For $r=3,4$, both $P_{n,r}$ and $P_{{\rm a},r}$ have unit high-frequency gain, so $\widetilde\gamma=1$ gives $\mu_\gamma=0$; set $\rho=1.2$. For $r=2$, neither $\rho$ nor $\varphi_p$ is used.}} Initialize all states at zero and use $w_r=0.2\sin(0.5t)$ and $d_r=0.15(1-e^{-2t})$. Let $R_1=0.8+0.2e^{-0.3t}$ and $R_c=1.6+0.4e^{-0.3t}$, and, for $r=3,4$, let $R_p=R_1/1.2$. {{Use $(\varphi_1,\phi)=(R_1^{-1},R_c^{-1})$ for $r=2$ and $(\varphi_1,\varphi_p,\phi)=(R_1^{-1},R_p^{-1},R_c^{-1})$ for $r=3,4$.}}

{On $\mathcal T_h:=\{0,0.01,\ldots,5\}$, let $\kappa_{\rm dom}^{(r)}$ be the largest sampled absolute value of $q_i$, $1\leq i<r$, and $\mathcal R_k(\eta^z_{1:k})$, $1\leq k\leq r$. The pairs $(\kappa_{\rm dom}^{(r)},\max_{\mathcal T_h}|v|)$ for $r=2,3,4$ are $(0.068,0.069)$, $(0.165,0.169)$, and $(0.888,4.218)$. Thus all sampled coordinates remain inside their domains; the $r=4$ case has the largest sampled correction peak.}

\end{document}